\documentclass[10pt,twocolumn,letterpaper]{article}

\usepackage[pagenumbers]{cvpr} 

\usepackage{times}
\usepackage{epsfig}
\usepackage{graphicx}
\usepackage{amsmath}
\usepackage{amssymb}

\usepackage{cite}
\usepackage{amsmath,amssymb,amsfonts}
\usepackage{graphicx}
\usepackage{textcomp}
\usepackage{xcolor}

\usepackage{url}
\usepackage{algorithm,algorithmic}
\usepackage{booktabs,makecell,multirow}
\usepackage[accsupp]{axessibility}

\usepackage[pagebackref,breaklinks,colorlinks]{hyperref}

\usepackage[capitalize]{cleveref}
\crefname{section}{Sec.}{Secs.}
\Crefname{section}{Section}{Sections}
\Crefname{table}{Table}{Tables}
\crefname{table}{Tab.}{Tabs.}

\def\cvprPaperID{*****} 
\def\confName{CVPR}
\def\confYear{2023}

\begin{document}

\title{Multimodal Continuous Emotion Recognition: A Technical Report for ABAW5}

\author{Su Zhang\textsuperscript{1}, Ziyuan Zhao\textsuperscript{1,2}, Cuntai Guan\textsuperscript{1, \thanks{This work is partially supported by the RIE2020 AME Programmatic Fund, Singapore (No. A20G8b0102).}} \\
\textsuperscript{1}Nanyang Technological University\\
\textsuperscript{2}Institute of Infocomm Research (I$^2$R), A*STAR, Singapore\\
{\tt\small sorazcn@gmail.com, S210088@e.ntu.edu.sg, ctguan@ntu.edu.sg}
}
\maketitle

\begin{abstract}
We used two multimodal models for continuous valence-arousal recognition using visual, audio, and linguistic information. The first model is the same as we used in ABAW2 and ABAW3, which employs the leader-follower attention. The second model has the same architecture for spatial and temporal encoding. As for the fusion block, it employs a compact and straightforward channel attention, borrowed from the End2You toolkit. Unlike our previous attempts that use Vggish feature directly as the audio feature, this time we feed the pre-trained VGG model using logmel-spectrogram and finetune it during the training. To make full use of the data and alleviate over-fitting, cross-validation is carried out. The code is available at https://github.com/sucv/ABAW3.
\end{abstract}

\begin{figure*}[!htbp]
\centering
\includegraphics[width=\textwidth]{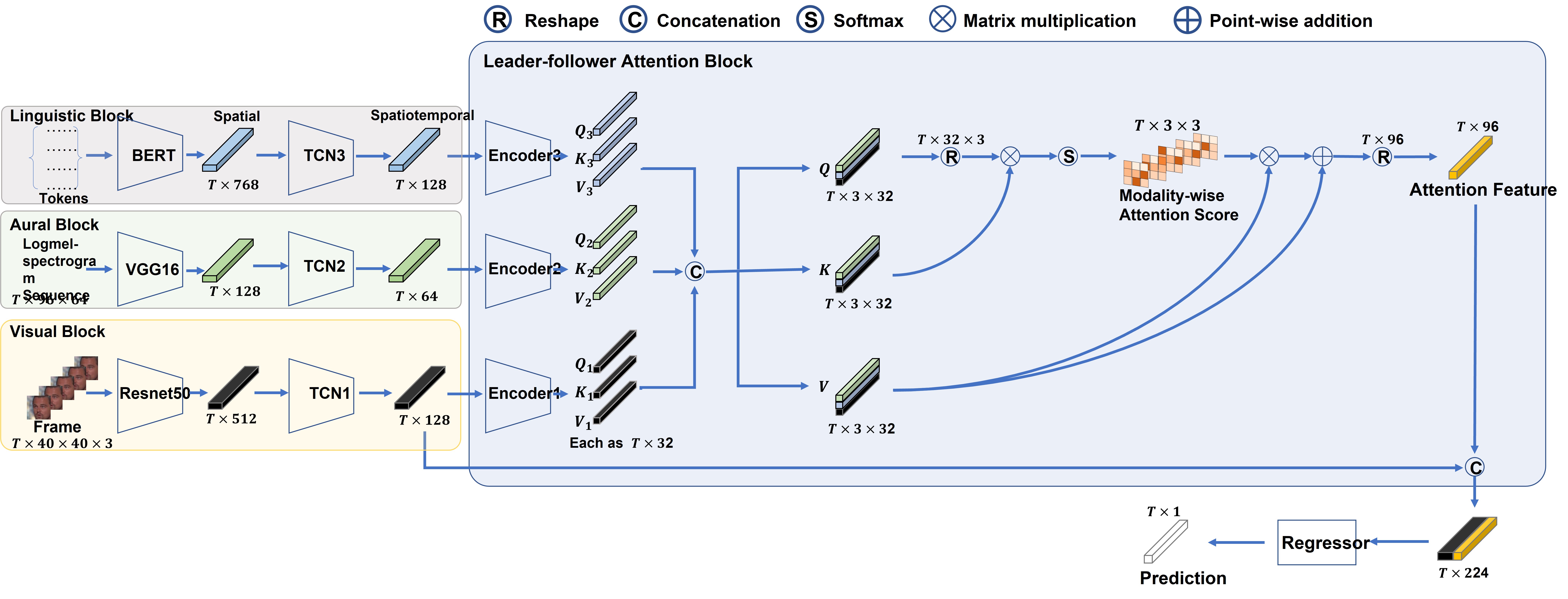}
\caption{The architecture of our proposed model. The model consists of four components, i.e., the visual, audio, linguistic, and co-attention blocks. The visual block has a cascaded 2DCNN-TCN structure, and the audio and linguistic blocks each contain a TCN. The three branches yield three independent spatial-temporal feature vectors. They are then fed to the attentive fusion block. Three independent attention encoders are used. For the $i$-th branch, its encoder consists of three independent linear layers, they adjust the dimension of the feature vector producing a query $\mathbf{Q}_i$, a key $\mathbf{K}_i$, and a value $\mathbf{V}_i$. They are then regrouped and concatenated to form the cross-modal counterparts.}\label{fig:model}
\end{figure*}

\section{Introduction}
Emotion recognition is the process of identifying human emotion. It plays a crucial role in behavioral modeling, human-computer interaction, and affective computing. By using the dimensional model \cite{sandbach2012static}, any emotional state can be taken as a point located in a continuous space, with the valence and arousal being the axes. Continuous emotion recognition seeks to map the $N$ sequential data points into $M$ sequential emotional state points, where $M$ usually equals $N$. The term "continuous" is endowed with two distinctive features. Spatially, it seeks to represent the emotional state as a point with continuous numerical values in the multi-dimensional space of the dimensional theory, as opposed to assigning categorical labels. Temporally, it incessantly forecasts the emotional state for a predetermined duration, forming a continuous record of the subject's emotional trajectory over a specified period.

The task of Continuous Emotion Recognition (CER) is challenging, mainly due to the following reasons. Firstly, the complexity of emotions over an extended period can be seen as a composition of multiple facial muscular movements, which can be defined as one-actions \cite{hussein2019timeception} using the Facial Action Coding System (FACS) \cite{sankaran2020domain}. However, despite its atomic nature, human emotions, irrespective of the modality, typically exhibit large variations in their intensity, order, and duration, and take longer to unfold. Consequently, there is a need for models that can learn long-range temporal dependencies to address this issue. Secondly, emotions are highly subjective and vary significantly depending on individual experiences, making their perception susceptible to personal bias. For instance, children who have been subjected to physical abuse can detect signs of anger much faster than their counterparts \cite{myers2004psychology}. Therefore, the data collected from subjects and the ground truth provided by annotators are prone to individual biases. One promising technique to mitigate this issue is multimodality, which involves the integration of visual, audio, and physiological data to develop reliable CER models.

In order to address the challenges posed by temporal dynamic learning and cross-subject generality, we adopt two strategies: large window resampling and multimodal fusion. Large window resampling involves selecting a window size that determines the amount of contextual information the model considers to make a prediction for each time step. On the other hand, multimodal fusion is a widely accepted technique that leverages the complementary nature of multiple data modalities to improve prediction accuracy. This is because multimodal data can disambiguate conflicting emotional cues, such as a crying face with joyful vocal expressions, or a neutral face with a harsh intonation. Additionally, multimodal data can enhance recognition robustness in real-world scenarios characterized by uncontrolled environments, subject diversity, various illumination conditions, spontaneous behaviors, background noise, and unclear speech. For instance, when an actor's face is not visible, their voice or body gesture can serve as an emotional cue to support ongoing learning.

This report details our methodology for the valence-arousal estimation challenge from the fifth affective behavior analysis in-the-wild (ABAW5) workshop \cite{kollias2022abaw, kollias2021analysing, kollias2020analysing, kollias2021distribution, kollias2021affect, kollias2019expression, kollias2019face, kollias2019deep, zafeiriou2017aff}. ABAW5 aims for the affective behavior analysis in-the-wild. The valence-arousal estimation sub-challenge bases on the extended version of the Aff-Wild2 database, in which the valence and arousal are annotated continuously for each video frame. Our work is an extension of the last year's attempt \cite{zhang2022continuous} on ABAW5 \cite{kollias2021analysing}. The extension are twofold. First, a pre-trained audio VGGnet is employed to serve as the audio backbone for the vggish \cite{hershey2017cnn} extraction on-the-go. Thus, the expressiveness of deep feature on visual and audio modalities can be further improved through finetuning. Second, a compact and straightforward fusion block is borrowed from the End2You toolkit, in which the channel attention is employed to fuse the multimodal information. The results of using leader-follower attention and channel attention are reported.




The remainder of the paper is arranged as follows. Section \ref{sec:model_architecture} details the model architecture including the visual, aural, linguistic, and attentive fusion blocks. Section \ref{sec:implementation_details} elaborates the implementation details including the data pre-processing, training settings, and post-processing. Section \ref{sec:result} provides the continuous emotion recognition results. Section \ref{sec:conclusion} concludes the work.

\section{Model Architecture}
\label{sec:model_architecture}
There are two architectures for our model, the leader-follower attention network (LFAN) and channel attention network (CAN). Both of them consists of independent branches to extract the spatiotemporal encoding for each modality, followed by the fusion block using leader-follower attention and channel attention, respectively. For each branch, a CNN backbone is employed to extract the spatial deep feature on-the-go for video frames and logmel-spectrogram, followed by a temporal convolutional network (TCN) \cite{bai2018empirical} to further learn the spatiotemporal encoding. For the linguistic branch, the pre-extracted bert features \cite{radford2019language} are taken as the input to feed the TCN. For branches fed by low-level features, no backbone is employed as well. The illustration of LFAN using video frames, logmel-spectrograms, and speech tokens are shown in Fig \ref{fig:model}. As for the details of the channel attention-based fusion block, please refer to the End2You paper \cite{tzirakis2018end2you} (\href{https://github.com/end2you/end2you/blob/master/end2you/models/multimodal/fusion/fusion_layer.py}{code}).

\section{Implementation Details}
\label{sec:implementation_details}
\subsection{Database}
The ABAW5 competition uses the Aff-Wild2 database. The corpora of the valence-arousal estimation sub-challenge includes $564$ trials. The database is split into the training, validation and test sets. The partitioning is done in a subject independent manner so that any subject's data are included in only one partition. The partitioning produces $341$, $71$, and $152$ trials for the training, validation, and test sets. Four experts annotate the videos using the method proposed in \cite{cowie2000feeltrace}. In addition to the annotations (for the training and validation sets only) and the raw videos, the bounding boxes and landmarks for each raw video are also available to the participants.

\subsection{Preprocessing}
The visual preprocessing is carried out as follows. The cropped-aligned image data provided by the organizer are used. All the images are resized to $48\times 48\times 3$ as a compromise to limited computational power. Given a trial from the training or validation set, the length $N$ is determined by the number of the rows of the annotation text file which does not include $-5$. For the test set, the length $N$ is determined by the frame number of the raw video. A zero matrix $\mathbf{B}$ of size $N\times 48\times 48\times 3$ is initialized and then iterated over the rows. For the $i$-th row of $\mathbf{B}$, it is assigned as the $i$-th jpg image if it exists, otherwise doing nothing. 

The audio preprocessing firstly converts all the videos to mono with a $16K$ sampling rate in wav format. The logmel-spectrogram are then extracted using the preprocessing code from the Vggish repository\footnote{https://github.com/harritaylor/torchvggish}. The only change is that we specified the hop length to be $1/fps$ of the raw video, in order to synchronize with other modalities and annotations.  

The linguistic preprocessing is carried out as follows. The mono wav file obtained from the audio preprocessing is fed to a pretrained speech recognition model from the Vosk toolkit\footnote{https://alphacephei.com/vosk/models/vosk-model-en-us-0.22.zip}, from which the recognized words and the word-level timestamp are obtained. The recognized words are then fed to a pretrained punctuation model from Hugging Face\footnote{https://huggingface.co/felflare/bert-restore-punctuation}. After which a pretrained BERT model from the Pytorch library is employed to extract the word-level linguistic features. The linguistic features are obtained by summing together the last four layers of the BERT model \cite{sun2020multi}. To synchronize, the word-level linguistic features are populated according to the timestamp of each word and each frame. Specifically, a word usually has a larger time span than that for a frame. Therefore, for one word, its feature is repetitively assigned to the time steps of all the frames within the time span.

For the valence-arousal labels, all the rows containing $-5$ are excluded. To ensure that the features have the same length as the corresponding trial, the feature matrices are either repeatedly padded (using the last feature points) or trimmed (starting from the rear), depending on whether the feature length is shorter than the trial length or not, respectively.

\begin{table*}[ht]
\centering
\caption{The CCC results from the 6-fold cross-validation on the validation set. Fold 0 is exactly the original data partitioning provided by ABAW5.}\label{table:result}
\setlength{\tabcolsep}{4mm}{
\begin{tabular}{cccccccc}
\toprule
 
\makecell{Emotion} &\makecell{Method}&\makecell{ Fold 0}& \makecell {{\textbf{Fold 1}}}  &\makecell{Fold 2} &\makecell{Fold 3} &\makecell{Fold 4} &\makecell{ Fold 5}       \\
 
\midrule
\multirowcell{2}{Valence}&LFAN&$0.441$&$-$&$-$&$-$&$-$&$-$\\
&CAN&$0.423$&$\mathbf{0.612}$&$0.484$&$0.529$&$0.543$&$0.488$\\
\midrule
\multirowcell{2}{Arousal}&LFAN&$0.645$&$-$&$-$&$-$&$-$&$-$\\
&CAN&$0.670$&$\mathbf{0.680}$&$0.592$&$0.618$&$0.655$&$0.610$\\

\bottomrule
\end{tabular}}
\end{table*}

The audio preprocessing firstly converts all the videos to mono with a $16K$ sampling rate in wav format. After which, the logmel-spectrogram is extracted following the Vggish preprocessing code \footnote{https://github.com/harritaylor/torchvggish}. Note that the hop length for these features are set to be $1/fps$ of the raw video, in order to synchronize with other modalities and annotations.  

The linguistic preprocessing is carried out as follows. The mono wav file obtained from the audio preprocessing is fed to a pretrained speech recognition model from the Vosk toolkit\footnote{https://alphacephei.com/vosk/models/vosk-model-en-us-0.22.zip}, from which the recognized words and the word-level timestamp are obtained. The recognized words are then fed to a pretrained punctuation and capitalization model from the deepmultilingualpunctuation toolkit\footnote{https://pypi.org/project/deepmultilingualpunctuation/}. After which a pretrained BERT model from the Pytorch library is employed to extract the word-level linguistic features. The linguistic features are obtained by summing together the last four layers of the BERT model \cite{sun2020multi}. To synchronize, the word-level linguistic features are populated according to the timestamp of each word and each frame. Specifically, a word usually has a larger time span than that for a frame. Therefore, for one word, its feature is repetitively assigned to the time steps of all the frames within the time span.

For the valence-arousal labels, all the rows containing $-5$ are excluded. To ensure that the features and annotations have the same length, the feature matrices are either repeatedly padded (using the last feature points) or trimmed, depending on whether the feature length is shorter and longer than the trial length, respectively.

\subsection{Data Expansion}

The AffWild2 database employed by ABAW5 contains $360$, $72$, and $162$ trials in the training, validation, and testing sets, respectively. To make full use of the available data and alleviate over-fitting, 6-fold cross-validation is employed. By evenly splitting the training set into $5$ folds, we have $6\times 72$ trials in total. Note that the $0$-th fold is exactly the original data partitioning. And there is no subject overlap across different folds. Moreover, during training and validation, the resampling window has a $33\%$ overlap, resulting in $33\%$ more data. 

\subsection{Training}
The batch size is $12$. For each batch, the resampling window length and hop length are $300$ and $200$, respectively. I.e., the dataloader loads consecutive $300$ feature points to form a minibatch, with a stride of $200$. For any trials having feature points smaller than the window length, zero padding is employed. For visual data, the random flip, random crop with a size of $40$ are employed for training and only the center crop is employed for validation. The data are then normalized so that $mean=std=0.5$. For bert features, they are normalized so that $mean=0$ and $std=1$.

The CCC loss is used as the loss function. The Adam optimizer with a weight decay of $0.001$ is employed. The maximal epoch number and early stopping counter are set to $100$ and $20$, respectively. The learning rate (LR) and minimal learning rate (MLR) are set to $1e-5$ and $1e-8$, respectively. 

The warmup scheme with \textit{ReduceLROnPlateau} scheduler with a patience of $5$ and factor of $0.1$ is employed based on the validation CCC. Three groups of layers for the visual Resnet50 backbone and audio VGG backbone are manually selected for further fine-tuning. When epoch$=0$, the first group is unfrozen. The learning rate is then linearly warmed up to $1e-5$ within an epoch. The repetitive warm-up is carried out until epoch$=5$. After which the \textit{ReduceLROnPlateau} takes over to update the learning rate. It gradually drops the learning rate in a factor of $0.1$ should no higher validation CCC appears for a consecutive 5 epochs. When the learning rate is already $1e-8$ and no higher validation CCC appears for the latest 5 epochs, the second group is unfrozen, the learning rate is also reset to $1e-5$ before undergoing the warm-up scheme and \textit{ReduceLROnPlateau} again. The procedure is repeated until no groups are available to unfrozen. Note that at the end of each epoch, the current best model state dictionary (i.e., the one with the greatest validation CCC) is loaded. By doing so, when there is no improvement on validation CCC, the training CCC cannot keep increasing and ends up with over-fitting. 

Note that unlike our previous attempts on ABAW2 and ABAW3, no post-processing is employed. Given the results from 6 folds, we simply choose the ones with the highest validation CCC to submit. Also note that the results for any fold are selected from $15$ instances using different seeds.

\section{Result and Discussion}
\label{sec:result}

The validation results of our methods against the baseline are reported in Table \ref{table:result}. Note that for one emotion type, the reported result for each fold is the best instance selected from 15 instances using different seeds. Finally, the best results of Fold 0 from CAN and LFAN, together with other 3 best folds (measured in the average CCC) from CAN, are submitted to ABAW5. 

In this study, two multimodal models are compared, namely CAN and LFAN, with the former utilizing only video frames and logmel-spectrograms while the latter incorporates an additional Bert feature. Validation results indicate that the CAN model with two modalities outperforms the LFAN model with three modalities, achieving a higher CCC score in Fold 0. It is worth noting that attempts to include the Bert feature as a third modality in the CAN model did not result in improvement. This suggests that the current linguistic information encoded by the pre-trained Bert model does not provide useful information to the model, or at least not in its current form. The issue lies in the fact that the length of tokens is not equivalent to the length of the sampling window, rendering fine-tuning impossible during training. Future research is needed to explore how linguistic information can be effectively incorporated into the fusion process. Furthermore, the study highlights the efficacy of fine-tuning. Both Resnet50 and VGG16 models, trained using large public datasets and fed by video frames and logmel-spectrograms, show improvements in performance when the last few layers are fine-tuned during training. This improvement can be as high as $10\%$ or more, compared to models that use pre-extracted CNN and Vggish features, albeit at a cost of $300\%-500\%$ longer training time.

The best results from all the teams are reported in Table \ref{table:all}. 

\begin{table}
\centering
\caption{The overall test results in CCC. The bold fonts indicate the best results. Citations for several teams are not available from Google Scholar by the time when this report was drafted.}\label{table:all}
\vspace{0.25cm}
\begin{tabular}{|c|c|c|c|}
\hline
Method                     & Valence & Arousal & Mean  \\ \hline
SituTech & 0.619 & \textbf{0.663} & \textbf{0.641}\\ \hline
Netease Fuxi Virtual Human \cite{zhang2023multimodal} & \textbf{0.649} & 0.626 & 0.637 \\ \hline
Ours  & 0.553 & 0.630 & 0.591 \\ \hline
CtyunAI \cite{zhou2023continuous} & 0.501 & 0.633 & 0.567 \\ \hline
HFUT-MAC \cite{zhang2023facial} & 0.523 & 0.545 & 0.534 \\ \hline
HSE-NN-SberAI \cite{savchenko2023emotieffnet}& 0.482 & 0.528 & 0.505 \\ \hline
ACCC & 0.462 & 0.506 & 0.482 \\ \hline
PRL \cite{vu2023vision} & 0.504 & 0.428 & 0.466 \\ \hline
SCLAB CNU \cite{nguyen2023transformerbased}  & 0.458 & 0.470 & 0.464 \\ \hline
USTC-AC \cite{wang2023facial} & 0.325 & 0.232 & 0.278 \\ \hline
Baseline \cite{kollias2023abaw}                  & 0.211             & 0.191             & 0.201  \\ \hline 
\end{tabular}
\end{table}



\section{Conclusion}
\label{sec:conclusion}
In our study, we aimed to develop effective models for continuous valence-arousal recognition, utilizing multimodal information from visual, audio, and linguistic sources. To overcome challenges posed by temporal dynamic learning and cross-subject generality, we employed two strategies: large window resampling and multimodal fusion. Specifically, we designed the CAN and LFAN models, with a resampling window length and hop length of $300$ and $200$, respectively, and multimodal fusion block to effectively integrate information from different modalities. Moving forward, our research will explore the incorporation of transformer-based networks for improved temporal dynamics learning.




{\small

\bibliographystyle{ieee_fullname}
\bibliography{egbib}
}

\end{document}